\documentclass[preprint,11pt,axodraw]{JHEP3}
\usepackage{axodraw,epsfig}
\usepackage{subfigure}

\newcommand\lsim{\mathrel{\rlap{\lower4pt\hbox{\hskip1pt$\sim$}}
    \raise1pt\hbox{$<$}}}
\newcommand\gsim{\mathrel{\rlap{\lower4pt\hbox{\hskip1pt$\sim$}}
    \raise1pt\hbox{$>$}}}
\def\bea{\begin{eqnarray}}
\def\eea{\end{eqnarray}}
\def\ba{\begin{array}}
\def\ea{\end{array}}
\def\bc{\begin{center}}
\def\ec{\end{center}}
\def\nn{\nonumber}

\def\la{\langle}
\def\ra{\rangle}

\def\p{\prime}
\def\f{\frac}
\def\a{\alpha}
\def\d{\dagger}

\def\g{\gamma}

\def\L{\Lambda}
\def\f#1#2{\frac{#1}{#2}}

 \preprint{SNUTP 06-008}
\title{Mediation of Supersymmetry Breaking \\
in Gauge Messenger Models}
\author{Radovan Derm\' \i\v sek$^*$, Hyung Do Kim$^\dagger$ and Ian-Woo Kim$^\dagger$\\
$^*$School of Natural Sciences, Institute for Advanced Study,
Princeton, NJ 08540, U.S.A.\\
$^\dagger$School of Physics and Astronomy and Center for Theoretical Physics,\\
Seoul National University, \\
Seoul, 151-747, Korea\\
E-mail: \email{dermisek@ias.edu, hdkim@phya.snu.ac.kr, iwkim@phya.snu.ac.kr}}

\abstract{ We study gauge mediation of supersymmetry breaking in
SU(5) supersymmetric grand unified theory with gauge fields as
messengers. The generated soft supersymmetry breaking parameters
lead to close to maximal mixing scenario for the Higgs mass and
highly reduce the fine tuning of electroweak symmetry breaking. All
gaugino, squark and slepton masses are determined by one parameter
-- the supersymmetry breaking scale. The characteristic features
are: negative and non-universal squark and slepton masses squared at
the unification scale, non-universal gaugino masses, and sizable
soft-trilinear couplings. In this scenario, all soft supersymmetry
breaking parameters at the unification scale can  be smaller than
400 GeV and all the superpartners can be lighter than 400 GeV and
still satisfy all the limits from direct searches for superpartners
and also the limit on the Higgs mass. The lightest supersymmetric
particle is gravitino or a sizable mixture of bino, wino and
higgsino. We also consider a possible contributions from additional
messengers in vector-like representations, and a contribution from
gravity mediation, which is estimated to be comparable. }

\begin{document}

\section{Introduction}

Minimal Supersymmetric Standard Model (MSSM) is one of the most
promising candidates for physics beyond the standard model. Gauge
coupling unification, radiative electroweak symmetry breaking (EWSB)
and the lightest supersymmetric particle (LSP) as a candidate for
dark matter in the presence of R-parity indicate that MSSM might be
the correct description of physics above the EW scale.

A natural explanation of EWSB being triggered by SUSY breaking
requires the SUSY breaking scale to be near the EW scale. However,
we have not observed any superparticles yet. Moreover, the Higgs
quartic coupling in the MSSM is solely determined by gauge
couplings, which gives a definite prediction for the physical Higgs
mass. At tree level it is lower than the Z boson mass ($M_Z \simeq
91$ GeV),
\bea m_h & \le & M_Z |\cos 2\beta|, \label{eq:mh_tree} \eea where
$\tan \beta =
v_u/v_d$ is the ratio of the vacuum expectation values (VEVs) of
$H_u$ and $H_d$. The dominant one loop correction
\cite{Okada:1990vk} \cite{Haber:1990aw} \cite{Ellis:1990nz} , in
case the stop mixing parameter is small, depends only
logarithmically on stop masses and it has to be large in order to
push the Higgs mass above the LEP limit, 114.4 GeV. A two loop
calculation (we use {\it FeynHiggs
2.4.1}~\cite{Heinemeyer:1998yj,Heinemeyer:1998np} with $m_t = 172.5$
GeV) reveals the stop masses have to be $\gsim 900$ GeV.

This constraint has a direct drawback in the electroweak symmetry
breaking. The mass of the Z boson (or the EW scale), determined by
minimizing the Higgs potential, is related to the supersymmetric
Higgs mass parameter $\mu$ and the soft SUSY breaking mass squared
parameter for $H_u$ as (for $\tan \beta \ge 5$) \bea \f{M_Z^2}{2} &
\simeq & -\mu^2 (M_Z) - m_{H_u}^2 (M_Z). \label{eq:MZ} \eea The
large stop mass affects the running of $m_{H_u}^2$, \bea \delta
m_{H_u}^2 & \simeq & -\f{3}{4\pi^2} m_{\tilde{t}}^2 \log
\f{\Lambda}{m_{\tilde{t}}}. \label{eq:del_mhu} \eea and, since for
$\Lambda \sim M_{\rm GUT} \sim 10^{16}$ GeV the loop suppression
times large log is of order one, we find
\bea \delta m_{H_u}^2 & \sim & - m_{\tilde{t}}^2. \nn \eea
Comparing it with Eq.~(\ref{eq:MZ}) we immediately see that we need
a miraculous cancelation between $m_{H_u}^2$ and $\mu^2$ to obtain
the right $M_Z$ for $m_{\tilde{t}} \gsim 900$ GeV. One possibility
to keep $\mu$ of order $M_Z$ is to start with large enough
$m_{H_u}^2$ at the GUT scale to cancel the large log correction
$-m_{\tilde{t}}^2$ in which case the fine tuning is hidden in the
boundary condition for $m_{H_u}^2$. This is the so called ``little
hierarchy problem".

The situation highly improves when considering large mixing in the
stop sector. The mixing is controlled by the ratio of $A_t - \mu
\cot \beta$ and $m_{\tilde{t}}$, where $A_t$ is the soft SUSY
breaking top trilinear coupling. Since we consider parameter space
where $\mu$ is small to avoid fine tuning and $\tan \beta \gsim 5$
in order to maximize the tree level Higgs mass~(\ref{eq:mh_tree}),
the mixing is simply given by $A_t / m_{\tilde{t}}$. The Higgs mass
is maximized for $A_t(M_Z) / m_{\tilde{t}} (M_Z) \simeq \pm
\sqrt{6}$ and with such a mixing the limit on the Higgs mass can be
satisfied  with much lower stop masses, $m_{\tilde{t}} (M_Z)
\lesssim 300$ GeV. Therefore in this ``maximal mixing scenario"
(scenario where mixing in the stop sector is such that the Higgs
mass is maximized) the fine tuning in EWSB is highly alleviated.
However it is very non-trivial to realize this scenario in models,
since it usually requires very large $A_t$ at the GUT scale, several
times larger than other soft SUSY breaking parameters. The maximal
mixing scenario and its possible realization in models will be
discussed in more detail in Sec.~\ref{sec:max_mixing}.

A simple way of achieving close to maximal mixing was recently
suggested in \cite{Dermisek:2006ey}. If we allow negative stop
masses squared at the GUT scale several interesting things happen
simultaneously. First of all, unless $m_{\tilde{t}}$ is too large
compared to $M_3$ it will run to positive values at the EW scale. At
the same time the contribution to $m_{H_u}^2$ from the energy
interval where $m_{\tilde{t}}^2 <0 $ partially or even exactly
cancels the contribution from the energy interval where
$m_{\tilde{t}}^2 >0 $, see Eq.~(\ref{eq:del_mhu}), and so the EW
scale value of $m_{H_u}^2$ can be arbitrarily close to the starting
value at $M_{GUT}$. No cancelation between initial value of
$m_{H_u}^2$ (or $\mu$) and the contribution from the running is
required. And finally, the stop mixing is typically much larger than
in the case with positive stop masses squared. It turns out that in
the region where $m_{H_u}^2$ gets negligible contribution from
running, the radiatively generated stop mixing is close to maximal
even when starting with negligible mixing at the GUT scale. Since in
principle this scenario can eliminate fine tuning of EWSB
completely, it is desirable to see how close to the radiatively
generated maximal mixing scenario one can get in specific models.

In this paper we study gauge mediation of SUSY breaking in SU(5)
supersymmetric grand unified theory (SUSY GUT) with an adjoint
chiral multiplet and massive components of vector (gauge) multiplet
playing the role of messengers. The soft susy breaking parameters in
this ``gauge messenger model" are similar to those discussed in
\cite{Dermisek:2006ey} which were shown to lead to maximal mixing
scenario for the Higgs mass. The characteristic features are:
negative and non-universal squark and slepton masses squared at the
GUT scale, non-universal gaugino masses, $|M_1|  >  |M_2| > |M_3|$,
and sizable soft-trilinear couplings. Besides gauge messengers, we
also consider a possible contributions from additional messengers in
vector-like representations, e.g. $5$ and $\bar{5}$ of SU(5).
Finally, since the messenger scale is the GUT scale, and the gauge
mediation is a one loop effect, the naively estimated size of
gravity mediation induced by non-renormalizable operators
(suppressed by $M_{\rm Pl}$) is comparable to the contribution from
gauge mediation. A combination of gauge mediation (with gauge and
vector-like messengers) with gravity mediation opens completely new
possibilities for model building. We show that already some of the
simplest models lead to close to maximal mixing scenario for the
Higgs mass and highly reduce the fine tuning of electroweak symmetry
breaking. The SUSY spectrum is very different from other scenarios
typically used for collider studies. All superpartners can be within
400 GeV with relatively light stop, $m_{{\tilde t}_1}\gsim 150$,
while satisfying all experimental limits, including the limit on the
Higgs mass. The lightest supersymmetric particle (LSP) is gravitino
and the next to the lightest supersymmetric particle (NLSP) 
is neutralino, sneutrino, stau or stop
in most of the parameter space.

We note that gauge messenger model has been considered in very early
stages of MSSM history. After the work on inverted mass
hierarchy~\cite{Witten:1981kv}, ``geometric hierarchy model" has
been constructed in~\cite{Dimopoulos:1982gm} and soft scalar masses
have been calculated in \cite{Kaplunovsky:1983yx}.{\footnote{See
also more recent attempts to break GUT symmetry and SUSY by the same
field in \cite{Agashe:1998kg} \cite{Agashe:2000ay}.}} In this model
the SUSY breaking scale is an intermediate scale and the messenger
scale is the GUT scale. The explicit SUSY breaking model they
considered has light (TeV scale) adjoint chiral superfields under
the standard model gauge group and the gauge couplings unify at a
scale beyond the Planck scale, $10^{20}$ GeV. We do not consider a
specific model of SUSY breaking (although we assume it happens at
the GUT scale). We only address the mediation of SUSY breaking.
Therefore, we treat the number of fields in a model as discrete
parameters and focus on minimal models with smaller number of
fields.

This paper is organized as follows. In Sec.~\ref{sec:max_mixing} we
discuss the maximal mixing scenario as a possible solution to the
little hierarchy problem, and a possibility of it being generated
radiatively without introducing large soft-trilinear couplings at
the GUT scale. In Sec.~\ref{sec:GMM} we present a gauge messenger
model and briefly discuss possible contribution from gravity mediation
of SUSY breaking.
The results are given
in Sec.~\ref{sec:results} together with discussion of phenomenology.
We conclude in Sec.~\ref{sec:conclusions}. For convenience we summarize formulae
necessary to derive soft SUSY breaking parameters from gauge
messenger models in the Appendix~\ref{sec:softterms}, and we discuss
different possibilities for gravity mediated contributions in more detail in the
Appendix~\ref{sec:gravity}.

\section{Maximal mixing scenario -- a solution to the fine tuning problem \label{sec:max_mixing}}

As mentioned in the Introduction, the physical Higgs boson mass
receives an additional contribution from stop mixing
\cite{Ellis:1991zd}, \bea m_h^2 & \simeq & M_Z^2 \cos^2 2\beta +
\f{3G_F m_t^4}{\sqrt{2} \pi^2} \left\{ \log
\f{m_{\tilde{t}}^2}{m_t^2} + \f{A_t^2}{m_{\tilde{t}}^2}
(1-\f{A_t^2}{12 m_{\tilde{t}}^2} ) \right\}. \label{eq:mh_mix} \eea
The last term has a maximum at $|A_t/m_{\tilde{t}}| = \sqrt{6}$
which corresponds to the maximal mixing scenario. In this case the
stop can be lighter, $m_{\tilde{t}}$ (maximal mixing) $ = e^{-3/2}
m_{\tilde{t}}$ (no mixing), and it can be as light as 250 $\sim$ 300
GeV while fulfilling the physical Higgs mass bound from the LEP.

Instead of using Eq.~(\ref{eq:del_mhu}) as a rough estimate of the
contribution of stop mass to the running of $m_{H_u}^2$ it is
instructive to be more precise. For given $\tan \beta$ we can solve
RG equations exactly and express EW values of $m_{H_u}^2$,
$\mu^2$, and consequently $M_Z^2$ given by Eq. (\ref{eq:MZ}), as
functions of all GUT scale parameters. For $\tan \beta =10$, we
have:
\bea M_Z^2 & \simeq & -1.9 \mu^2 + 5.9 M_3^2 -1.2 m_{H_u}^2 + 1.5 m_{\tilde{t}}^2
  - 0.8 A_t M_3 + 0.2 A_t^2 + \cdots,
\label{eq:MZ_gut} \eea where parameters appearing on the right-hand
side are the GUT scale parameters, we do not write the scale
explicitly. Other scalar masses and $M_1$ and $M_2$ appear with
negligible coefficients and we neglect them in our discussion. The
coefficients in this expression depend only on $\tan \beta$ (they do
not change dramatically when varying $\tan \beta$ between 5 and 50)
and $\log (M_{GUT}/M_Z)$.

Let us also express the EW scale values of stop mass squared, gluino
mass and top trilinear coupling for $\tan \beta = 10$ in a similar
way: \bea
m_{\tilde{t}}^2 (M_Z) & \simeq & 5.0 M_3^2 + 0.6 m_{\tilde{t}}^2   + 0.2 A_t M_3 \label{eq:mstop_gut} \\
M_3 (M_Z) & \simeq & 3 M_3 \label{eq:M3_gut} \\
A_t (M_Z) & \simeq & - 2.3 M_3 + 0.2 A_t. \label{eq:At_gut} \eea In
the case of $m_{\tilde{t}}$ the coefficients represent averages of
exact coefficients that would appear in separate expressions for $
m_{\tilde{t}_L}^2$ and $m_{\tilde{t}_R}^2$.

From Eqs.~(\ref{eq:MZ_gut}), (\ref{eq:mstop_gut}) and
(\ref{eq:M3_gut}), we see the usual expectation from SUSY, $M_Z
\simeq m_{\tilde{t}_{1,2}} \simeq m_{\tilde{g}}$, when all the soft
SUSY breaking parameters are comparable. Furthermore, neglecting
terms proportional to $A_t$ in Eqs.~(\ref{eq:At_gut}) and
(\ref{eq:mstop_gut}) we find that a typical stop mixing is \bea
\left|\f{A_t}{m_{\tilde{t}}} \right| (M_Z)  & \simeq & \f{2.3
M_3}{\sqrt{5.0 M_3^2 + 0.6 m_{\tilde{t}}^2}} \lesssim 1.0,
\label{eq:Atovermstop} \eea and comparing it with
Eqs.~(\ref{eq:mh_mix}) we see that such a mixing only negligibly
affects the mass of the Higgs boson. Due to the washout effect, see
Eq.~(\ref{eq:At_gut}), a large mixing can be achieved only for
$|A_t| \gg |M_3|, m_{\tilde{t}}$ at the GUT scale for opposite sign
of $A_t$ compared to $M_3$, or even larger $A_t$ for the same
sign.\footnote{Extremely large $A_t$ in the case of the same sign as
$M_3$ contributes significantly to the running of $m_{\tilde{t}}$
and consequently to the running of $m_{H_u}$. Therefore, $A_t \simeq
m_{\tilde{t}} \simeq m_{H_u} \gg M_Z$ is required and the EW scale
is a result of large cancelations.} This is the reason why it is
very difficult to build a model leading to the maximal mixing
scenario.

Although the boundary condition for $m_{\tilde{t}}$ in the above
discussion does not seem to be very important (it is mostly the
gluino that drives the evolution of stop and thus $m_{H_u}^2$, and
sets the mixing) it turns out that when considering negative stop
masses squared it starts playing a major role as discussed recently
in Ref.~\cite{Dermisek:2006ey}. In spite of negative stop masses
squared being somewhat suspicious, from Eq.~(\ref{eq:mstop_gut}) we
see that unless $m_{\tilde{t}}$ is too large compared to $M_3$ it
will run to positive values at the EW scale. At the same time,
however, the contribution to $m_{H_u}^2$ from the energy interval
where $m_{\tilde{t}}^2 <0 $ partially or even exactly cancels the
contribution from the energy interval where $m_{\tilde{t}}^2 >0 $
and so the EW scale value of $m_{H_u}^2$ can be arbitrarily close to
the starting value at $M_{GUT}$. From Eq.~(\ref{eq:MZ_gut}) we see
that this happens for $m_{\tilde{t}}^2 \simeq - 4 M_3^2$ (neglecting
$A_t$). No cancelation between initial value of $m_{H_u}^2$ (or
$\mu$) and the contribution from the running is required, the
electroweak scale is not sensitive to masses of colored particles in
this case, and the situation when $M_Z \ll m_{\tilde{t}_{1,2}}
\simeq m_{\tilde{g}}$ can be achieved without any fine tuning
(provided there exists a model which generates negative stop masses
squared and sets the ratio of gluino mass and the stop mass
approximately to the required value). And finally, from
Eqs.~(\ref{eq:mstop_gut}) and  (\ref{eq:At_gut}), or from
Eq.~(\ref{eq:Atovermstop}), we see that the stop mixing is typically
quite large. Unlike in the case with positive stop masses squared
where mixing is typically less than one, in the case with negative
stop masses squared it is typically greater than one, and it can be
easily even maximal. The maximal mixing scenario
 can be entirely generated radiatively starting
with no mixing at the GUT scale.

Very large $A_t$ term may cause dangerous color and/or charge
breaking minimum to appear at around the EW vacuum. Considering
cosmology, in order not to tunnel within the age of universe, the
empirical bound is~\cite{Kusenko:1996jn} \cite{Kusenko:1996xt} \bea
|A_t|^2 (M_Z)+ 3 \mu^2 (M_Z) \lesssim 7.5 (m_{\tilde{t}_L}^2 (M_Z) +
m_{\tilde{t}_R}^2 (M_Z)), \eea which is much weaker than the
condition for the EW vacuum to be the global minimum, $|A_t|^2 (M_Z)
+ 3\mu^2 (M_Z) \lesssim 3(m_{\tilde{t}_L}^2 (M_Z) +
m_{\tilde{t}_R}^2 (M_Z))$ \cite{Casas:1995pd}. Certainly the maximal
mixing value is within the empirical bound and it is safe from the
constraints of the CCB minima.

\subsection{Large (maximal) mixing in models}

Since the radiatively
generated maximal mixing scenario can in
principle eliminate fine tuning of EWSB
completely, it is desirable to see
whether it is possible to get even close to it
in specific models.

It is easy to see that this solution does not exist in mSUGRA. As a
consequence of universalities in gaugino and scalar masses, when
stop mass squared is negative enough to generate maximal stop mixing
at the EW scale radiatively, sleptons remain tachyonic even at the
EW scale because the bino contribution to the running of slepton
masses is small. The EW scale slepton mass is $m_{\tilde{e}_R}^2
\simeq m_0^2 + 0.15 M_{1/2}^2$ \cite{Feng:2005ba}. Imposing the
slepton mass bound 100 GeV gives the following inequality \bea m_0^2
& \ge & \left\{ -(0.4)^2 + (\f{100~{\rm GeV}}{M_{1/2}})^2 \right\}
M_{1/2}^2 . \eea The largest (negative) ratio of $m_0^2$ and
$M_{1/2}^2$ is achieved in the limit $M_{1/2} \to \infty$ (taking
aside all the naturalness criteria) and even in this case it is only
$m_0^2 \simeq -(0.4)^2 M_{1/2}^2$ which makes negligible difference
in the generated mixing at the EW scale, see
Eq.~(\ref{eq:Atovermstop}). The maximal mixing solution can be
achieved only when either gaugino masses are not universal at the
GUT scale (bino should be heavier than gluino at the GUT scale) or
scalar masses are not universal (sleptons are less negative than
stops).

Usual gauge mediation \cite{Dine:1993yw} \cite{Dine:1994vc}
\cite{Dine:1995ag}
shares a common problem with mSUGRA due to its
hierarchical spectrum at the weak scale. Gluino is almost 6 $\sim$ 7
times heavier than bino and squarks are much heavier than sleptons.
Anomaly mediation \cite{Randall:1998uk} \cite{Giudice:1998xp}
also has a huge hierarchy in the EW scale
spectrum and gluino is 10 times heavier than wino.

Recently proposed ``mirage mediation" or ``modulus-anomaly mixed
mediation" 
\cite{Choi:2004sx} \cite{Choi:2005ge} \cite{Choi:2005uz} \cite{Endo:2005uy}
\cite{Falkowski:2005ck} \cite{Choi:2005hd} \cite{Kitano:2005wc}
\cite{Lebedev:2005ge} partially fulfills the criteria listed above.
In the most interesting $\alpha =2$ scenario of mirage mediation
\cite{Choi:2005uz} \cite{Choi:2005hd} \cite{Kitano:2005wc}, the
mirage scale is at TeV and the spectrum is more or less degenerate.
In this case, squarks and sleptons are tachyonic except stop and
$H_u$ at the GUT scale and gaugino masses are non-universal at the
GUT scale with the aid of anomaly mediation. The fine tuning in this
model is highly reduced due to cancelation of RG running effects
with anomaly mediation contribution. The stop mixing is predicted to
be large but not close to the maximal, $|A_t/m_{\tilde{t}}| \sim
1.4$. The $\alpha = 2$ mirage mediation might be an alternative
solution to the little hierarchy problem although the supersymmetry
spectrum (except Higgs) can be at around TeV which is $4\pi$ times
heavier than $M_Z$. There are several common features between mirage
mediation and gauge messenger model considered in this paper though
the origin of supersymmetry breaking is very different.

In the next section we present a model of mediation of SUSY breaking
which leads to close to maximal mixing scenario while all the SUSY
breaking parameters at the GUT scale and also physical masses of
all superpartners can be $\lesssim 400$ GeV.

\section{Gauge Messenger Model \label{sec:GMM}}

Let us consider $N=1$ SU(5) supersymmetric grand unified theory (SUSY
GUT). The $N=1$ vector multiplet $V$ transforms as an adjoint
of SU(5), the three generations of matter fields are in chiral
multiplets, $3 \times (10+\bar{5})$, and the Higgs fields are in $5 +\bar{5}$.
Besides these, we also introduce an adjoint chiral multiplet $\Sigma$,
and we assume that both its scalar component, which we also denote $\Sigma$,
and the auxiliary component, $F_{\Sigma}$, get vacuum expectation values.
The VEV of $F_{\Sigma}$ breaks SUSY and the SUSY breaking is communicated
to gauginos, squarks and sleptons through gauge interactions.
The massive components of the gauge multiplet $V$ and $\Sigma$ play the role of
messengers. This is the minimal field content we consider.
In this case, the beta function coefficient
of the unified gauge coupling is $b_G = 3$ and all soft SUSY breaking parameters
at the GUT scale are calculable in terms of $b_G$ and the unified gauge coupling.

It is also possible to extend the messenger sector and introduce,
for example, a pair of usual messenger fields $\Phi$ and $\Phi^c$ in $5$
and $\bar{5}$ representations of SU(5). Additional messengers also change the
beta function coefficient, $b_G = 3$ and the spectrum is in general given
in terms of the number of messengers, $N_{\rm mess}$, and $b_G$.

Therefore, in this scenario
the mediation of supersymmetry breaking is a combination of two
effects:
\begin{itemize}
\item Gauge messenger contribution:

X and Y gauge bosons and gauginos contribute to the soft
supersymmetry breaking terms. They become massive by the VEV of $\Sigma$
and gaugino masses get split due to $F_\Sigma$. Therefore, the
messenger scale is the GUT scale.
The ratio $| \f{F_{\Sigma}}{\Sigma} |$ governs the common overall scale
of soft SUSY breaking parameters given by gauge messengers.
For convenience, we introduce $M_{\rm SUSY}$ defined as:
\bea M_{\rm SUSY} & = & \f{\alpha_{\rm GUT}}{4\pi} \left|
\f{F_{\Sigma}}{\Sigma} \right|,
\label{eq:MSUSY}
\eea
which we use in expressions for all soft SUSY breaking parameters.

\item Matter messenger contribution:

If the additional  vector-like messengers $\Phi$ and $\Phi^c$ couple
to $\Sigma$, \bea W & = & \Phi \Sigma \Phi^c, \eea they also
contribute to the soft SUSY breaking terms.\footnote{In principle it
is possible to introduce an additional singlet superfield, whose $F$
component is non-zero, and which couples to the vector-like matter
messengers. However, we consider only the minimal version in which
$\Phi$ and $\Phi^c$ couple to the adjoint $\Sigma$ by which gauge
messengers got their mass splitting.} The matter messengers also
become massive by $\Sigma$ VEV and mass splitting is given by
$F_{\Sigma}$. The same $M_{\rm SUSY}$ governs the common overall
scale of soft SUSY breaking parameters given by the matter
messengers.

\end{itemize}

The soft SUSY breaking parameters at the GUT scale (messenger scale)
can be calculated by the powerful and convenient technique, so
called ``analytic continuation into superspace"
\cite{Giudice:1997ni} \cite{Arkani-Hamed:1998kj}. The results are
derived in the Appendix~\ref{sec:softterms}, here we only summarize them.

Gaugino masses at the GUT scale ($\alpha_i = \alpha_{\rm GUT}$) are
(Eq.~(\ref{eq:gaugino})): \bea M_i & = & \left[ -2(5-N_{C_i}) +
N_{\rm mess} \right] M_{\rm SUSY}, \nn \eea where $N_{C_i}$ is the
number of colors of the gauge group $SU(N_{C_i})$.
 More explicitly , \bea M_3 & = & (-4+ N_{\rm mess})
ˆM_{\rm SUSY}  \label{eq:M3}, \\
M_2 & = & (-6 + N_{\rm mess})
M_{\rm SUSY},  \\
M_1 & = & (-10 + N_{\rm mess}) M_{\rm SUSY}.  \label{eq:M1} \eea In
the minimal messenger model ($N_{\rm mess} =0$), the gaugino masses
at the messenger scale (the GUT scale) are \bea
M_3 & = & -4 M_{\rm SUSY},  \\
M_2 & = & -6 M_{\rm SUSY},  \\
M_1 & = & -10M_{\rm SUSY}.  \eea Note that $|M_1| > |M_2| >
|M_3|$ at the GUT scale ($|M_1| : |M_2| : |M_3| = 2.5 : 1.5 : 1$).
As a result of RG evolution, at the weak scale we find
$|M_1(M_Z)| : |M_2(M_Z)| : |M_3(M_Z)| \sim 1 : 1 : 2$.
This is quite different from scenarios with the universal gaugino masses at the GUT scale which
lead to gluino about 7 times heavier than bino at the EW scale.

Soft mass squared parameters for squarks and sleptons at the GUT
scale with $N_{\rm mess} =0$ are given as (see
Eq.~(\ref{eq:scalars})): \bea m_{\phi}^2 & = & \left( -2 \sum_i~
c_i~ b_{X_i} + 2 \Delta c_{\phi} ~ b_G \right) M_{\rm SUSY}^2, \eea
where $\Delta c = c_5 - \sum_{i=1}^3 c_i$ and $c_5, c_i$ are the
quadratic casimirs of $\phi$ field under $SU(5)$ and standard model
gauge groups, and $b_{X_i}$ are the contributions of messenger
fields to the beta function coefficient. Detailed expression is
given in the Appendix~\ref{sec:softterms}. When there are additional
chiral messengers, we would obtain (well known) additional gauge
mediation contribution \cite{Giudice:1998bp}. Explicit expressions
for squark and slepton masses at the GUT scale are given as : \bea
m_Q^2 & = & (-20 +3 b_G +\f{21}{10} N_{\rm mess} ) M_{\rm SUSY}^2,
\\
m_{u^c}^2 & = & (-16 + 4 b_G + \f{8}{5} N_{\rm mess} ) M_{\rm
SUSY}^2,
\\
m_{d^c}^2 & = & (-12 + 2 b_G + \f{7}{5} N_{\rm mess} ) M_{\rm
SUSY}^2,
\\
m_L^2 & = & (-12 + 3 b_G + \f{9}{10} N_{\rm mess} ) M_{\rm SUSY}^2,
\\
m_{e^c}^2 & = & (-12 + 6 b_G + \f{3}{5} N_{\rm mess} ) M_{\rm
SUSY}^2,
\\
m_{H_u,H_d}^2 & = & (-12 + 3 b_G + \f{9}{10} N_{\rm mess} ) M_{\rm
SUSY}^2.  š\eea In the minimal case ($N_{\rm mess} =
0$), expressions are simplified : \bea m_Q^2 & = & (-20 +3 b_G)
M_{\rm SUSY}^2,
\\
m_{u^c}^2 & = & (-16 + 4 b_G) M_{\rm SUSY}^2,
 \\
m_{d^c}^2 & = & (-12 + 2 b_G) M_{\rm SUSY}^2,
 \\
m_L^2 & = & (-12 + 3 b_G) M_{\rm SUSY}^2,
 \\
m_{e^c}^2 & = & (-12 + 6 b_G) M_{\rm SUSY}^2,
 \\
m_{H_u,H_d}^2 & = & (-12 + 3 b_G ) M_{\rm SUSY}^2.
š\eea

Soft tri-linear terms are also calculated by adding individual
contributions from three fields involved (Eq.~(\ref{eq:aterm})),
\bea A_{ijk} & = & A_i +
A_j + A_k, \\
A_{\phi_i} & = & 2 \Delta c_{\phi_i} M_{\rm SUSY}. \eea More
explicitly, \bea
A_t & = & 10 M_{\rm SUSY}, \\
A_b & = & 8 M_{\rm SUSY},  \\
A_{\tau} & = & 12 M_{\rm SUSY}. \eea The same result is given to the
first and the second generation soft tri-linear terms as it just
depends on gauge charges. Matter messengers ($N_{\rm mess} \neq 0$)
do not affect the boundary condition of soft tri-linear terms as in
usual gauge mediation.

Negative sign in gaugino masses is absorbed by $U(1)_R$ symmetry
rotation. $A$ and $\mu$ terms change sign accordingly. Thus, we
choose the convention of $M_3 > 0$ in which $A < 0$ for $N_{\rm
mess} \le 4$.

\subsection{Characteristic Features}

Gauge messenger models are very predictive, since the soft SUSY breaking
parameters are calculable in terms of $M_{\rm SUSY}$
and gauge quantum numbers of
fields involved. The pattern of soft SUSY breaking terms is unique and
distinctively different from other models. The most striking features are:
\begin{itemize}

\item Non-universal gaugino masses at the GUT scale:
\bea
M_1 & > & M_2 > M_3 \nn
\eea
The gaugino masses are non-universal even at the GUT scale though we
started from the GUT models. It is the most interesting feature of
the gauge messenger model. Furthermore, bino (and wino) is heavier
than gluino at the GUT scale and the three gauginos have a tendency of
gathering at the EW scale due to the usual running behavior of
gauge couplings.

\item (Non-universal) Negative squarks and sleptons masses squared at the GUT scale:

Gauge messenger contribution alone typically leads to the squarks
and sleptons tachyonic at the GUT scale. However, this does not rule
out the theory and just imply that we might live in a meta-stable
vacuum rather than the true vacuum. From the discussion of fine
tuning we learned that it actually might be more natural to live in
a meta-stable vacuum. For $0 \le b_G \le 3$, which is the case in
realistic models due to a non-minimal content, squarks are even more
negative, $|m_{\tilde{q}}^2| > |m_{\tilde{l}}^2|$, $m_{\tilde{q}}^2
< 0$, $m_{\tilde{l}}^2 < 0$.

\item Sizable $A$ -- terms:

Large $A$ -- terms is one of the unique feature of gauge
messenger models which is absent in the usual gauge mediation. In usual
gauge mediation, the soft tri-linear terms at the messenger scale are
zero and are generated only by RG running.
Here $A_t$ is sizable and it will help to achieve close to maximal mixing
scenario.

\end{itemize}

\subsection{Contribution from Gravity Mediation}

Since the messenger scale is the GUT scale, and the gauge mediation
is a one loop effect, the naively estimated size of gravity
mediation induced by non-renormalizable operators (suppressed by
$M_{\rm Pl}$) is comparable to the contribution from gauge
mediation. The typical scale of gauge mediation is $M_{\rm SUSY}$,
given in Eq.~(\ref{eq:MSUSY}), and the typical size of gravity
mediation is $ m_{3/2} = \left|\f{F}{\sqrt{3} M_{\rm Pl}}\right|$.
Gravity to gauge mediation ratio is then \bea \f{m_{3/2}}{M_{\rm
SUSY}} & = & \f{4\pi M_{\rm GUT}}{\sqrt{3} \alpha_{\rm GUT} M_{\rm
Pl}} \simeq 1.5. \nn \eea Taking into account group theoretical
factors appearing in the formulas for gauge mediation we see that
the contribution of gravity mediation is of order 20\% or 30\% of
gauge mediation for order one coupling of non-renormalizable
operators.

There are several ways to deal with the contribution from gravity.
It is possible to suppress this contribution entirely, e.g. by
raising the cutoff scale of a theory beyond the Planck scale in
superconformal frame or by lowering the GUT scale. Alternatively,
one can actually use the contribution from gravity to generate the
$\mu$ term through the Giudice-Masiero mechanism
\cite{Giudice:1988yz}. The contribution from gravity can be also
made universal, or sector dependent. Different possibilities for
gravity contribution are discussed in detail in
Appendix~\ref{sec:gravity}.

A combination of gauge messengers with gravity mediation clearly opens an unexplored
direction for model building.
When we present results in the next section we take a pragmatic approach and consider only the
simplest possibilities for the contribution from gravity.

\section{Results: SUSY spectrum, the Higgs mass and the LSP \label{sec:results}}

In this section, we discuss SUSY and Higgs spectra in gauge
messenger models. SUSY spectrum is calculated with
SoftSusy~\cite{Allanach:2001kg} and for the calculation of the
lightest CP even Higgs mass we use {\it FeynHiggs
2.4.1}~\cite{Heinemeyer:1998yj,Heinemeyer:1998np} (with $m_t =
172.5$ GeV). We focus mainly on the minimal scenario of gauge
messenger model, $N_{\rm mess} =0$, $b_G = 3$, and only briefly
discuss other choices of $N_{\rm mess}$ and $b_G$. Depending on the
way gravity mediation contributes to the soft SUSY breaking
parameters we distinguish the following cases:

\begin{itemize}

\item Pure gauge mediation:

The model is the most predictive when we assume the gravity
contribution is suppressed to a negligible level. The suppression
does  not have to be huge since gauge mediation already dominates
over the gravity mediation. Given the particle content of a model
($N_{\rm mess}$ and $b_G$), a single parameter $M_{\rm SUSY}$
determines all the soft SUSY breaking parameters in terms of
measured gauge couplings and group theoretical factors. We do not
address the origin of $\mu$ and $B\mu$ terms in this case and we
treat them as free parameters (as usual, we exchange $B\mu$ for
$\tan \beta$).

Independent parameters : $M_{\rm SUSY}$, $\mu$ and $\tan \beta$.

\item Gauge mediation with gravity contribution in the Higgs sector:

In this case we consider that only the Higgs sector gets a sizable
contribution from gravity
mediation. This opens a possibility of generating the $\mu$ term
through  Giudice-Masiero mechanism. The soft masses squared of
$H_u$, $H_d$, and the $\mu$ and $B\mu$ terms are determined by $m_{3/2}$ with
order one couplings.

Independent parameters : $M_{\rm SUSY}$, $m_{H_u}^2$, $m_{H_d}^2$,
$\mu$ and $\tan \beta$.

\item Additional universal gravity contribution to scalar masses:

We also consider a possibility of adding
universal scalar masses to the two scenarios above. Adding
universal scalar masses does not change the spectrum in a crucial way (unless this contribution is huge).
However, small addition to scalar masses might change the LSP in some region
of parameter space, and consequently be responsible for very different phenomenology.

Additional independent parameters: $m_0$.

\end{itemize}

Finally, we also calculate fine tuning necessary for correct EWSB
\cite{Ellis:1986yg} \cite{Barbieri:1987fn}, defined as: \bea
\Delta_p \equiv \left| \frac{\partial \ln M_Z }{\partial \ln p}
\right|. \eea where $p$ spans over free parameters in a given model.
It can be easily estimated from the formula for $M_Z^2$,
Eq.~(\ref{eq:MZ_gut}), customized for a given case, e.g. in the case
of pure gauge mediation we have \bea M_Z^2 & \simeq & -1.9 \mu^2 +
\alpha  M_{\rm SUSY}^2, \eea where $\alpha$ depends on  $N_{\rm
mess}$, $b_G$ and $\tan \beta$. The fine tuning, $\Delta_\mu \simeq
\Delta_{M_{\rm SUSY}}$ in this case, gives us the precision with
which the two terms have to cancel each other.

\subsection{Pure Gauge Mediation}

\begin{figure}[t]
  \begin{center}
      \epsfig{figure=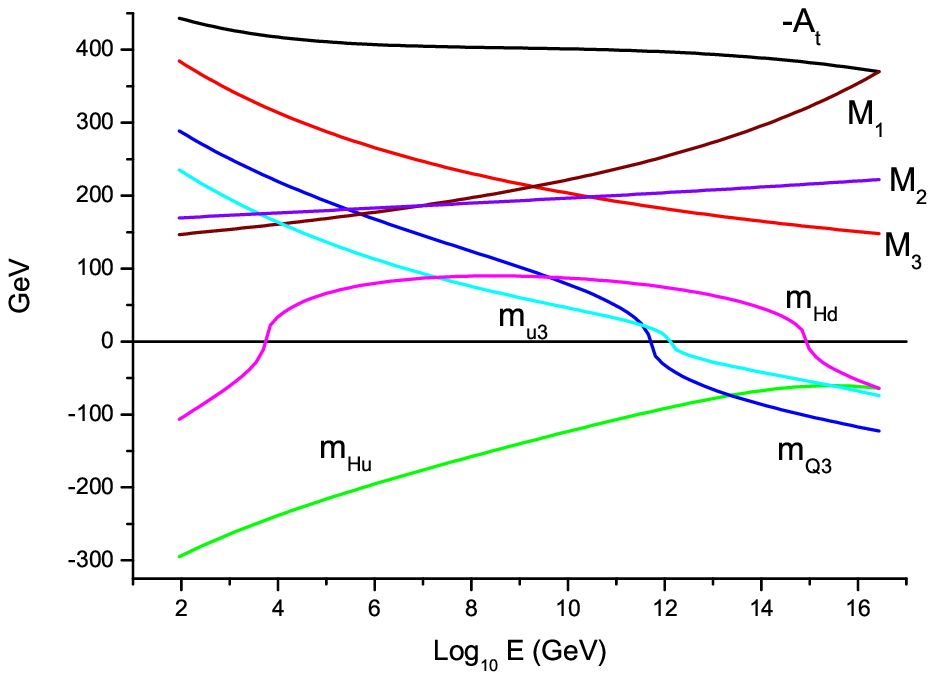,width=7.cm}
      \epsfig{figure=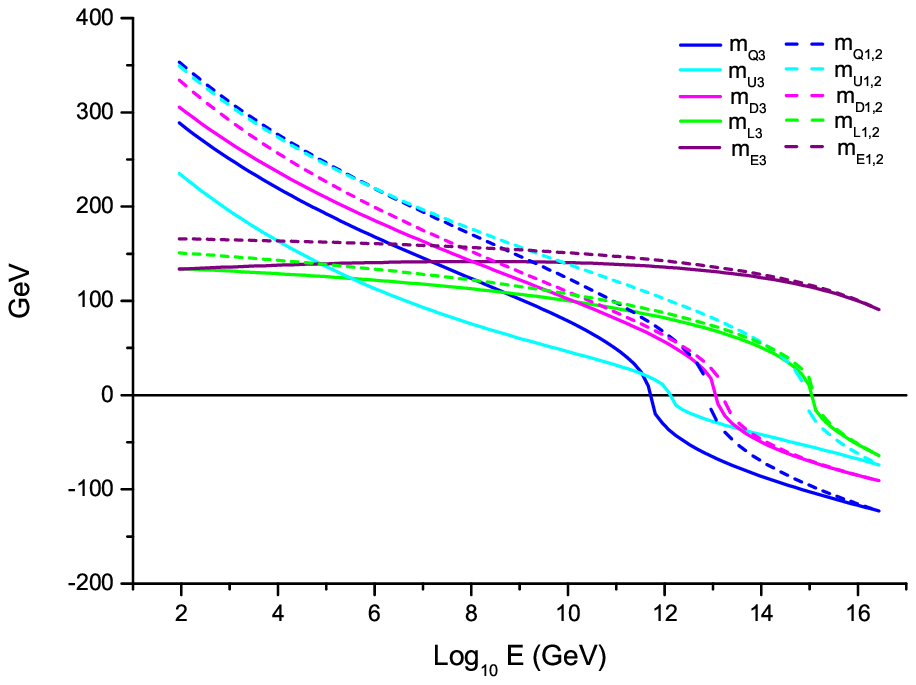,width=7.cm}
  \end{center}
\vspace{-0.5cm}
\caption{ \label{fig:G_RGrunning}
Renormalization group running of soft SUSY breaking parameters for pure gauge mediation,
$N_{\rm mess} =0$ and $b_G = 3$, with
$M_{\rm SUSY} = 37~{\rm GeV}$ and $\tan \beta = 23$. On the left:
evolution of gaugino masses, $A_t$, and stop and Higgs soft masses.
On the right: evolution of squark and slepton masses of the third generation (solid lines)
and the first two generations (dashed lines).
In order to have both mass dimension one and two parameters on the same plot
and keep information about signs, we define $m_{H_u} \equiv
m_{H_u}^2/\sqrt{|m_{H_u}^2|} $ and similarly for other scalar masses.
}
\end{figure}

Let us start with the case of pure gauge mediation, $N_{\rm mess}
=0$ and $b_G = 3$. The absolute value of $\mu$ is fixed by requiring
proper EWSB and so only the sign of $\mu$ can be chosen.\footnote{We
chose the positive sign of $\mu$ in all results to be in principle
consistent with $b \to s \gamma$.}
In
Fig.~\ref{fig:G_RGrunning} we plot renormalization group running of
soft SUSY breaking parameters for a particular choice of $M_{\rm
SUSY}$ and $\tan \beta$ which leads to some of the lightest SUSY
spectrum possible given the current experimental bound on SUSY and
Higgs particles. The detailed information about this point is given
in the first column of Table~\ref{tab:points}. Varying $\tan \beta$
does not qualitatively change results and increasing $M_{\rm SUSY}$
scales the whole spectrum up.

The plot in Fig.~\ref{fig:G_RGrunning} is unlike anything we are familiar with from other
models of SUSY breaking.
None of the soft SUSY breaking parameters at the GUT scale is larger than 400 GeV and
none of the superpartner is heavier than 400 GeV, and yet all the limits
from direct searches for SUSY particles and also the limit on the Higgs mass are satisfied.
\begin{figure}[t]
\begin{center}
\epsfig{figure=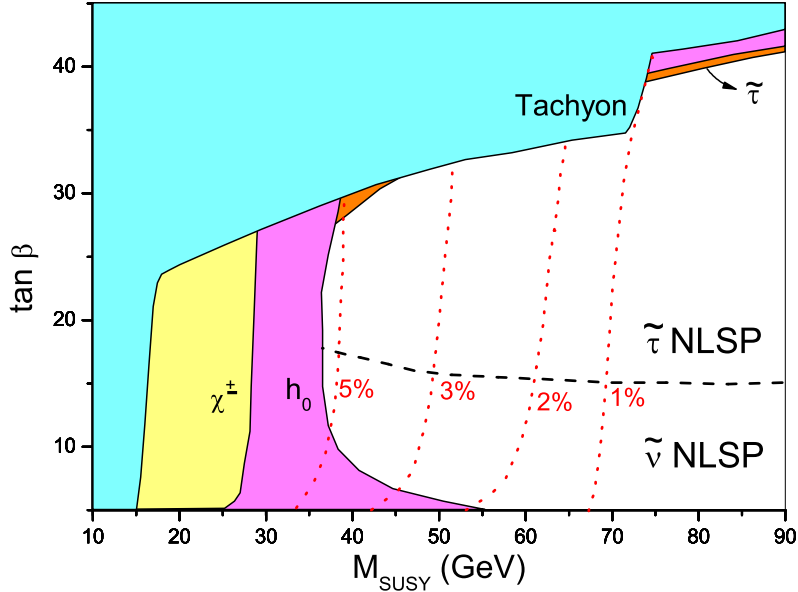, width=12cm}
\end{center}
\vspace{-1.cm}
\caption{ \label{fig:Gspace}
Allowed region of parameter space and the degree of fine tuning in
the $M_{\rm SUSY}-\tan \beta$ plane for pure gauge mediation,
$N_{\rm mess} =0$ and $b_G = 3$.
The shaded regions are excluded by direct searches for SUSY and Higgs particles.
We use the limits on the mass of the
lightest Higgs boson, $m_{h_0} > 114.4~{\rm GeV}$, the lightest stop,
$m_{\tilde{t}} > 95.7~{\rm GeV}$, the lightest stau,
$m_{\tilde{\tau}} > 81.9~{\rm GeV}$, and the lightest chargino,
$m_{\chi^{\pm}} > 117~{\rm GeV}$.
The region denoted as ``tachyon" is  excluded  due to tachyonic spectrum.
The black dashed line separates regions where sneutrino or stau is NLSP.
}
\end{figure}
Squark and slepton masses squared start negative at the GUT scale
(except right-handed sleptons, in this case)
and are driven to positive values by gaugino masses.
First two generations of squarks and sleptons are somewhat heavier
as in scenarios starting with positive scalar masses squared at the GUT scale.
Gluino is much lighter than in most models as a result of the hierarchical
boundary condition at the GUT scale, $|M_1| > |M_2| > |M_3|$.
The soft trilinear coupling, $A_t$ is sizable at the GUT scale,
which helps to achieve close to maximal mixing scenario.
On the other hand, sizable $A_t$ also contributes to the running
of $m_{H_u}^2$ proportional to $-|A_t|^2$, see Eq.~(\ref{eq:MZ_gut}).
The smallest possible $\mu$ in this case is about 270 GeV which require about
$5 \%$ tuning between $\mu$ and $M_{\rm SUSY}$ to recover the correct $M_Z$.\footnote{The current
limit on chargino mass requires $\mu \gsim 150$ GeV. Thus any model which
does not relate the $\mu$ term
in a calculable way to soft SUSY breaking parameters requires at least 20 \%  tuning from $\mu$.}

Since there are only two parameters in this model, it is easy to
explore the whole parameter space. In Fig.~\ref{fig:Gspace} we show
allowed parameter space in $M_{\rm SUSY}$ -- $\tan \beta$ plane,
together with regions excluded by direct searches for SUSY and Higgs
particles. Moderate to large $\tan \beta$ is allowed and, as usual,
as small $M_{\rm SUSY}$ which still satisfies the limit on the Higgs
mass is preferred by naturalness. In most region of the parameter
space sneutrino is NLSP for small tan beta (gravitino is the LSP)
and stau is NSLP for large tan beta 
due to large mixing of the left and right-handed stau.
A representative point from this
region is given in the first (stau NLSP) and the third column (sneutrino NLSP) 
of Table~\ref{tab:points}. 
As we will discuss later, small contributions from gravity mediation can
easily push sneutrino or stau above the lightest neutralino leading to
a large region where neutralino is (N)LSP.

\subsection{Gauge mediation with gravity contribution in the Higgs sector}

\begin{figure}[t]
  \begin{center}
      \epsfig{figure=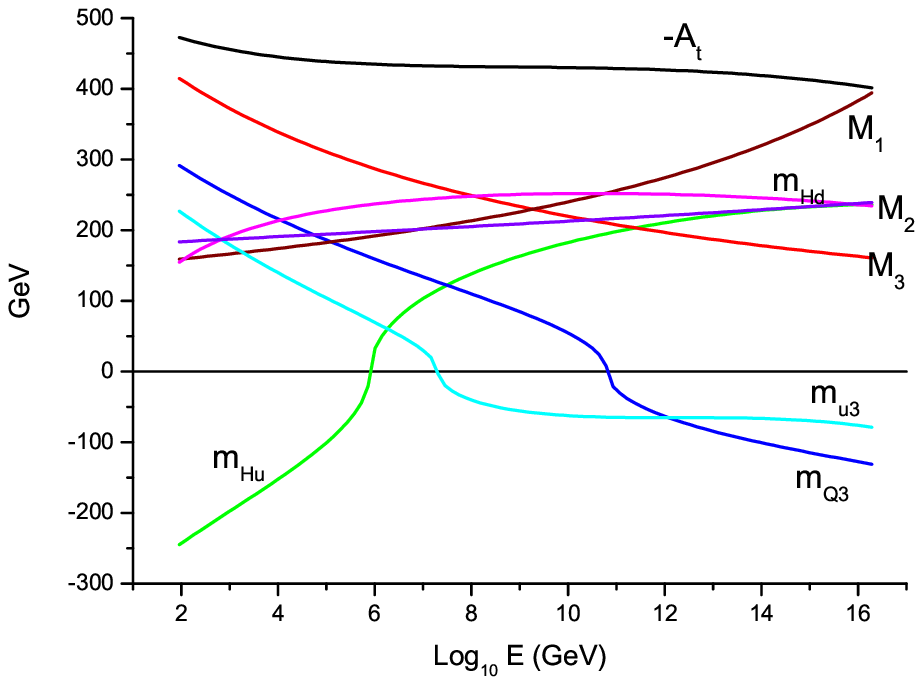,width=7.cm}
      \epsfig{figure=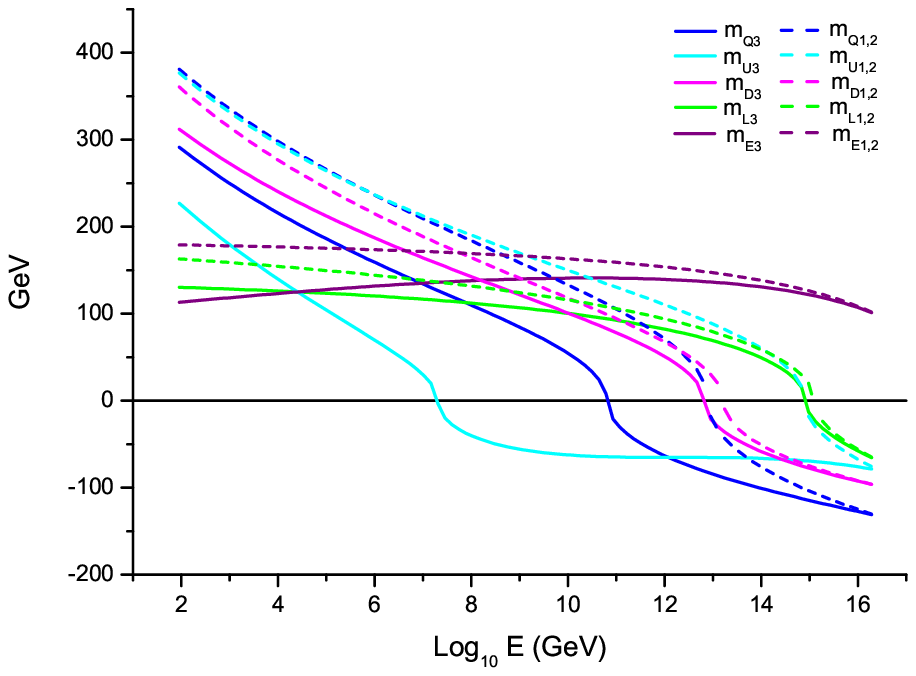,width=7.cm}
  \end{center}
\vspace{-0.5cm}
\caption{ \label{fig:GH_RGrunning}
Renormalization group running of soft SUSY breaking parameters for gauge mediation,
$N_{\rm mess} =0$ and $b_G = 3$,
with a contribution from gravity mediation in the Higgs sector
for $M_{\rm SUSY}=40~{\rm GeV}$,
$\tan \beta = 29$, $c_{H_u}= 38$ and $c_{H_d} = 37$. The meaning of the lines is the same as
in Fig. 1.} 
\end{figure}

Adding a contribution from gravity mediation opens a possibility of
generating the $\mu$ term using Giudice-Masiero mechanism.
Comparable in size soft masses squared for $H_u$ and $H_d$ are also
generated. We parameterize additional contribution to the Higgs soft
masses squared by: $c_{H_u} M_{\rm SUSY}^2$ and $c_{H_d} M_{\rm
SUSY}^2$. An example of the renormalization group running of soft
SUSY breaking parameters in this case is given in
Fig.~\ref{fig:GH_RGrunning} and detailed information about this
scenario can be found in the second column of
Table~\ref{tab:points}. Adding gravity contribution to soft Higgs
masses squared does not significantly affect running of other soft
SUSY breaking parameters. The major advantage of adding a positive
contribution to $m_{H_u}^2$ is that it allows smaller $\mu$ term.
This further reduces fine tuning of EWSB, see the
Table~\ref{tab:points}, because the original (somewhat large)
contribution from gauge mediation can be canceled in an equal way by
the additional contribution from gravity and by the $\mu$ term.

\begin{figure}[t]
\begin{center}
\epsfig{figure=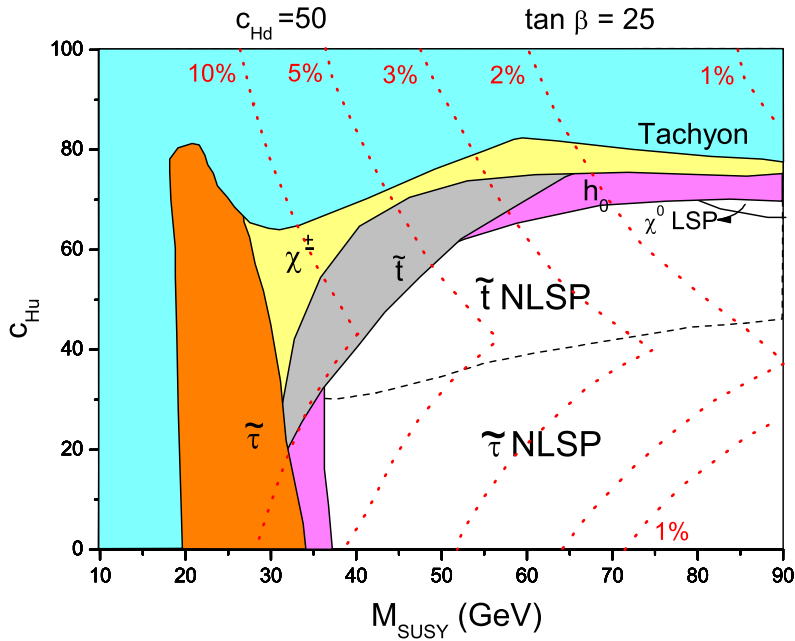, width=12cm}
\end{center}
\vspace{-1.cm}
\caption{ \label{fig:GHspace}
Allowed region of parameter space and the degree of fine tuning in
the $M_{\rm SUSY}-c_{H_u}$ plane for gauge mediation, $N_{\rm mess} =0$ and $b_G = 3$,
with a contribution from gravity mediation in the Higgs sector.
We fix $\tan \beta = 25$ and $c_{H_d} = 50.0$.
The meaning of excluded regions is the same as in Fig. 2. 
The black dashed line separates regions where ${\tilde{t}}_1$
and ${\tilde{\tau}}_1$ are NLSP.}
\end{figure}

Exploring the whole parameter space in this case is more complicated. In Fig.~\ref{fig:GHspace}
we present  a typical cut through the parameter space
in $M_{\rm SUSY} - c_{H_u}$ plane  with fixed $\tan \beta$ and $c_{H_d}$. We see that, depending
on the size of $c_{H_u}$, fine tuning from any of the parameters can be reduced to the level of 10\%.
Besides excluded regions that already appeared in the case of pure gauge mediation, Fig.~\ref{fig:Gspace},
there is also a region excluded by limits on the stop mass. This is due to a subtle effect of larger
$m_{H_u}$ in the evolution of stop masses squared. Stop masses squared run to slightly smaller values
which increases stop mixing and consequently leads to much smaller value of the lightest stop
mass eigenstate.
For the same reason, besides neutralino (N)LSP and stau NLSP, there is a region with
stop NLSP. The NLSP situation can be easily changed when considering contribution from
gravity mediation also to squarks and sleptons.
%

\subsection{Other cases}

Adding a universal contribution to all scalar masses from gravity
mediation has a negligible effect on the EW scale value of
$m_{H_u}^2$. This can be easily seen from Eq.~(\ref{eq:MZ_gut}) in
which the terms containing $m_{H_u}^2$ and $m_{\tilde{t}}^2$
approximately cancel each other for $m_{H_u}^2 = m_{\tilde{t}}^2 =
m_0^2$ at the GUT scale. Therefore, adding $m_0$ (unless it is very
large) does not change fine tuning of EWSB. The contribution from
$m_0$ also makes stops heavier and reduces the mixing. This reduces
the Higgs mass and so only small values of $m_0$ are allowed for
small $M_{\rm SUSY}$ -- the region we are interested in. Small $m_0$
is however sufficient to change the NLSP of a model. For smaller
$\tan \beta$ it can highly enlarge the region where neutralino is
(N)LSP instead of sneutrino or stop, and for larger $\tan \beta$ it can basically
eliminate the region where stop is NLSP.

\begin{figure}[t]
  \begin{center}
    \subfigure[$N_{\rm mess} = 0, \; b_G = 3$]{
      \epsfig{figure=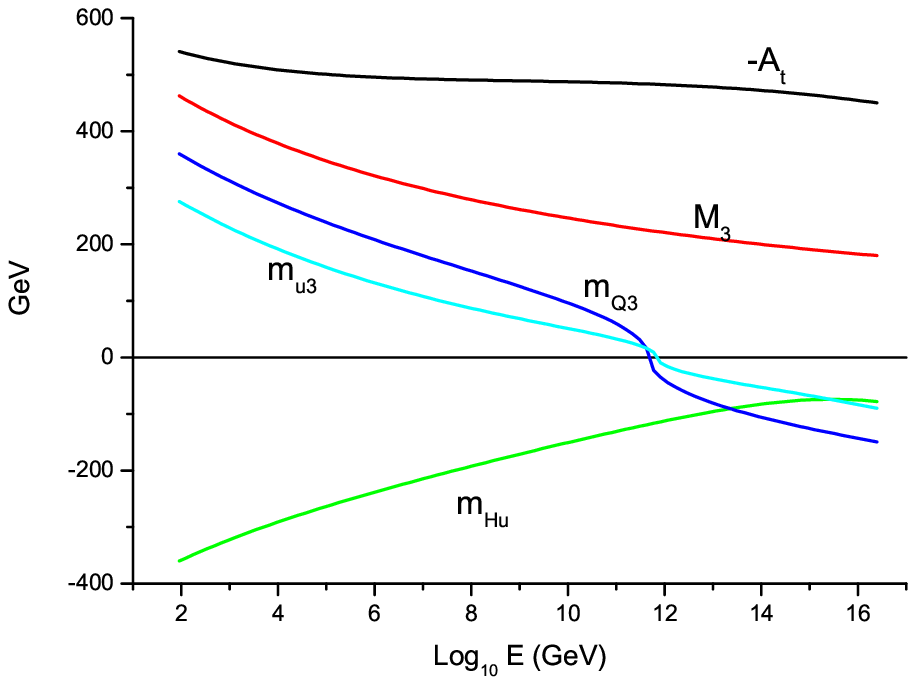,width=7cm}
    }
    \subfigure[$N_{\rm mess} = 0, \; b_G = 2$]{
      \epsfig{figure=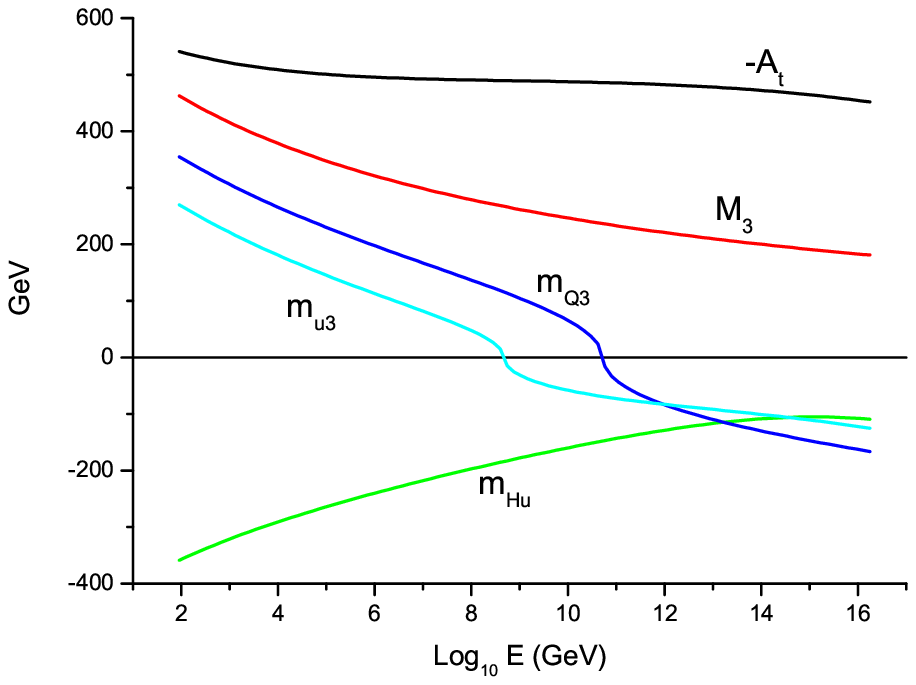,width=7cm}
    } \\
    \subfigure[$N_{\rm mess} = 0, \; b_G = 1$]{
      \epsfig{figure=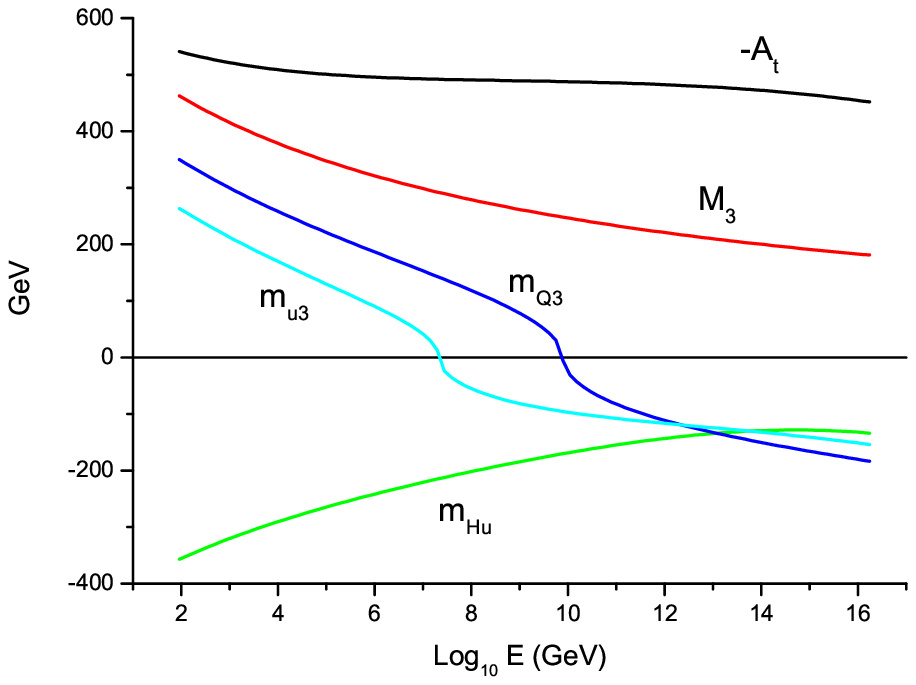,width=7cm}
    }
    \subfigure[$N_{\rm mess} = 0, \; b_G = 0$]{
      \epsfig{figure=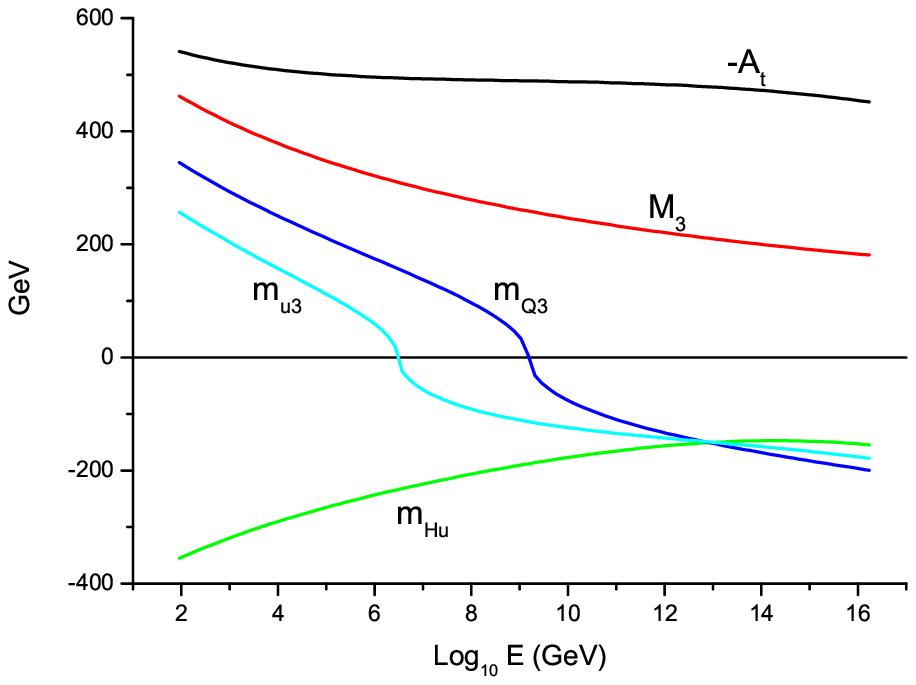,width=7cm}
    }
    \subfigure[$N_{\rm mess} = 1, \; b_G = 2$]{
      \epsfig{figure=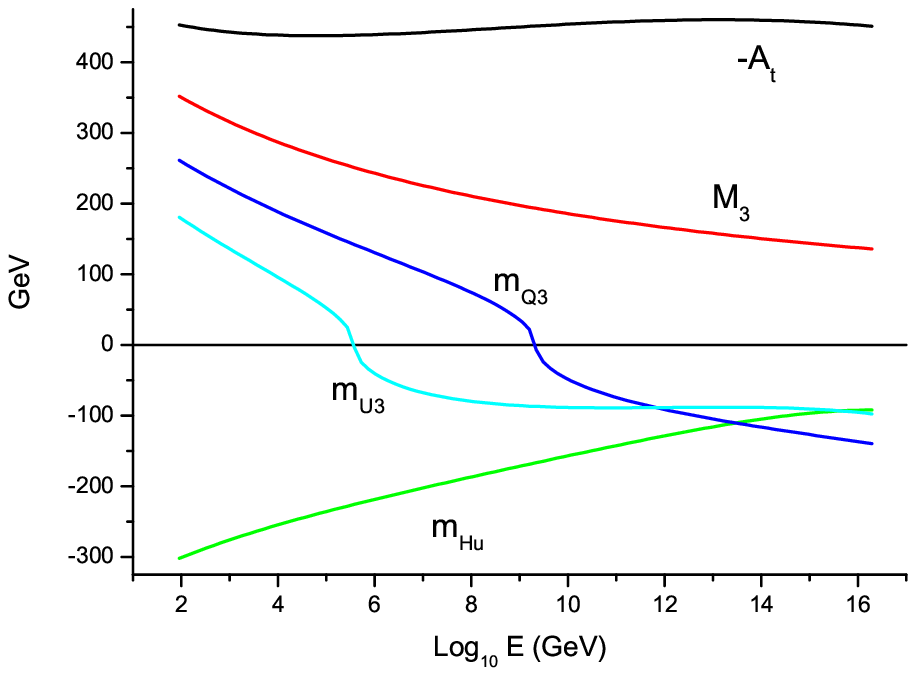,width=7cm}
    }
  \end{center}
\caption{ \label{fig:G_NbG_RGrunning}
Renormalization group running of relevant soft SUSY breaking parameters for pure gauge mediation,
for different choices of $N_{\rm mess}$ and $b_G$, with
$M_{\rm SUSY}= 45~\rm GeV$ and $\tan \beta = 8$.}
\end{figure}

So far we have discussed only the case with $N_{\rm mess} =0$ and $b_G = 3$.
In Fig.~\ref{fig:G_NbG_RGrunning} we
also present plots of  renormalization group running of soft SUSY breaking parameters
for $N_{\rm mess} = 0$ and smaller values of $b_G$ which correspond to adding more content to the
minimal GUT scenario. And for completeness, in the same figure we also include $N_{\rm mess} =
1$, $b_G = 2$ case which corresponds to the minimal GUT content with one pair of additional
vector-like messengers in $5$ and $\bar 5$ of SU(5). In all cases $M_{\rm SUSY}$ and $\tan \beta$
are fixed to the same value which allows us to see trends in the spectrum from changing the content
of a model. For exactly this reason we do not require that all the experimental limits are satisfied
in all models. Detailed information about these five points is given in the last five columns in
Table~\ref{tab:points}.
The basic features of all presented cases are very similar. Lowering
$b_G$ results in lighter squark and slepton spectrum but slightly
larger stop mixing. As a result, the Higgs mass is decreasing very
slowly. Adding additional pair of messenger leads to lighter
spectrum because of the cancelation between contributions from gauge
messengers and vector-like messengers and thus in order for this
scenario to be viable, larger $M_{\rm SUSY}$ is needed. We do not
discuss possible addition of gravity mediation for these scenarios.

\begin{table}[t]
  \begin{tabular}{|c||c|c||c|c|c|c|c|}
\hline 
$(N_{\rm mess}, b_G)$ & $(0,3)$ & $(0,3)$ & $(0,3)$ & $(0,2)$ & $(0,1)$ & $(0,0)$ & $(1,2)$ \\
\hline \hline
GUT parameter& & & & & & & \\
\hline
$M_{\rm SUSY}$ & 37  & 40 & 45 & 45 & 45 & 45 & 45\\
$\tan \beta$ & 23  & 29 & 8 &  8 & 8 & 8 & 8 \\
$c_{H_u}$     &     & 38 &   &    &   &  &  \\
$c_{H_d}$     &     & 37 &   &    &   &  &  \\
\hline
EW scale    & & & & & & & \\
parameter   & & & & & & & \\
\hline
$m_{Q3}$     & 274 & 277  & 338  & 334  &  329 & 324  & 254 \\
$m_{U3}$     & 219  & 211  & 254  & 248  &  242 & 236  & 173 \\
$m_{D3}$     & 290  & 299  & 373  & 369  &  365 & 360  & 292 \\
$m_{L3}$     & 133  & 130  & 178  & 161  &  141 & 118  & 138 \\
$m_{E3}$     & 135  & 115  & 196  & 163  &  120 & 49.3 & 154 \\
$M_1$       & 149  & 161  & 183  & 183  &  183 & 183  &  162 \\
$M_2$       & 171  & 185  & 208  & 208  &  208 & 208  &  173 \\
$M_3$       & 369  & 400  & 440  & 440  &  441 & 441  &  345 \\
$\mu$       & 270  & 210  & 336  & 335  &  334 & 333 & 283 \\
$m_{\tilde{t}}$ & 245 & 242  & 293  & 288  &  282 & 276 & 210 \\
$A_t/m_{\tilde{t}}$ &-1.78 &-1.93 &-1.81 &-1.85 &-1.89 &-1.93 &-2.15 \\
\hline
Physical & & & & & & & \\
spectrum & & & & & & & \\
\hline
$m_{h_0}$           & 114.4 & 115.6 & 115.3 & 115.2 & 114.9 & 114.1 & 110.1 \\
$m_{A}$                 & 248 & 265   &374   &365   &355  &345  & 306 \\
$\tilde{t}_1$            & 138  & 101  & 192   & 182  & 171 & 159 & 44.0 \\
$\tilde{b}_1$            & 263  & 266   & 350   & 345  & 339 & 334 & 258  \\
$\tilde{\tau}_1$         & 103  & 88.2  & 182   & 159  & 123 & 61.0 & 141  \\
$\tilde{\nu}_{\tau}$   & 118 & 114 & 168 & 149 & 128 & 102 & 123 \\
$\tilde{u}_1,\tilde{c}_1$  & 340  & 365   & 405   & 396  & 386 & 375 & 309  \\
$\tilde{d}_1,\tilde{s}_1$  & 328  & 352   & 390   & 385  & 380 & 374 & 298  \\
$\tilde{e}_1,\tilde{\mu}_1$& 158  & 169   & 188   & 172  & 135 & 78.7 & 166 \\
$\tilde{g}$            & 379  & 406   & 451   & 450  & 449 & 447 & 348 \\
$\chi^{\pm}_1$          & 158  & 152   & 204   & 203  & 203 & 203 & 165 \\
$\chi^0_1$             & 141  & 137   & 179   & 178  & 178 & 178 & 156 \\
$\Psi_{3/2}$           & $>55.5$ & $>60$ & $>67.5$ & $>67.5$ & $>67.5$ & $>67.5$ & $>67.5$ \\
\hline
Fine  & & & & & & &  \\
Tuning  & & & & & & &  \\
\hline
$\Delta_{M_{\rm SUSY}}$ & 17.6 & 10.6 & 26.4 & 26.2 & 26.1 & 26.0 & 18.8 \\
$\Delta_{\mu}$        & 18.2 & 11.1 & 30.6 & 30.4 & 30.2 & 30.0 & 21.9 \\
$\Delta_{c_{H_u}}$     &      & 8.96 &      &      &      &     & \\
$\Delta_{c_{H_d}}$     &      & 0.462 &     &      &      &     & \\
\hline
  \end{tabular}
\caption{\label{tab:points} GUT input parameters, EW scale parameters,
physical spectrum and fine tuning for gauge messenger models specified
by $N_{\rm mess}$ and $b_G$.
All the mass parameters are
understood in GeV units. Here $m_{\tilde{t}} = \sqrt{m_{\tilde{t}_1}
m_{\tilde{t}_2}}$.
We present only masses of the lightest squark and slepton in each generation. }
\end{table}

\subsection{LSP and NLSP}

When there is a sizable contribution to the Higgs soft parameters
from gravity mediation, neutralino can be LSP or NLSP
depending on the gravitino mass.
In this region, neutralino is a
sizable mixture of bino, wino and Higgsino. Sizable mixture of bino
with higgsino/wino can give the right amount of thermal relic
density for dark matter when $\mu$, $M_1$ and $M_2$ are of order 100
GeV. In addition, the cross section for the direct detection is
larger compared to bino LSP which gives a better chance to observe
it.

In most region of allowed parameter space, sneutrino/stau or stop 
is NLSP and the LSP
is the gravitino. 
Gravitino LSP scenario with the right-handed stau NLSP 
has been studied in the
framework of supergravity \cite{Feng:2003jw} \cite{Ellis:2003dn}
\cite{Feng:2004zu} \cite{Feng:2004mt} \cite{Ellis:2005ii} 
\cite{Steffen:2006hw}. 
The life time of the stau NLSP is from $10^{6}$ sec to $10^{10}$ sec and we
might be able to detect it using a stopper.  As we provide a
specific model in which the gravitino LSP is very plausible, we can
get a more concrete prediction on NLSP lifetime and gravitino relic
density. Similar analysis should be done for the stop NLSP.

The gauge messenger model considered here generally predicts a very
light stop, $m_{\tilde{t_1}} \gsim 150$ GeV in the least fine tuned
parameter space. The Fig.~\ref{fig:GHspace} shows that stop becomes
NLSP if $c_{H_u} \gtrsim 30$. Stop NLSP has been studied in
\cite{Chou:1999zb} in the framework of low scale gauge mediation.
When gravitino is very light, the decay of stop NLSP can happen
quickly, within a minute, and the search for a possible collider
signal can be done. If gravitino mass is at around the weak scale,
stop decays long after the big bang nucleosynthesis (BBN). Usually
decays of particles having hadronic channels destroy the successful
agreement of BBN
and such scenarios are not considered.\footnote{We
thank Michael Peskin for discussion on this point.} Nonetheless the detailed
analysis of stop decay after BBN should be done to clear up this
issue. If stop NLSP with weak scale gravitino mass is consistent
with BBN, more natural parameter space is allowed.

\section{Conclusions \label{sec:conclusions}}

In this paper we studied gauge mediation of supersymmetry breaking
in SU(5) SUSY GUT with heavy gauge fields as messengers. We were led
to consider this gauge messenger model by recently discussed
possibility of generating the maximal mixing scenario for the Higgs
mass radiatively~\cite{Dermisek:2006ey}. In the optimal scenario
colored particles do not contribute to the renormalization group
running of the $m_{H_u}^2$ which in principle can eliminate fine
tuning of EWSB. The gauge messenger model does not lead to the
optimal scenario (only close to it), since stop masses are not
negative enough at the GUT scale. However, it still highly reduces
the fine tuning of EWSB and has many interesting features.

In this scenario
negative scalar masses squared at the GUT scale,
 with squarks more negative than sleptons, together with
non-universal gaugino masses, with $M_1> M_2 > M_3$, lead to a viable
spectrum at the EW scale in large portion of parameter space.
None of the soft SUSY breaking parameters at the GUT scale has to be larger than 400 GeV and
none of the superpartner has to be heavier than 400 GeV to satisfy
all the limits
from direct searches for SUSY particles and also the limit on the Higgs mass.
There is no other existing scenario with similar features. And yet,
just like anomaly mediation or the usual gauge mediation,
also this scenario
is governed by a single parameter - the SUSY breaking scale $M_{\rm SUSY}$.
The ratios of different soft SUSY breaking parameters are entirely
fixed by group theoretical factors. The main features of the spectrum do not
change when considering more complicated GUT models
than the minimal scenario we focused on in this paper.
And finally, considering contributions from gravity mediation not only opens a possibility to generate
the $\mu$ term through the Giudice-Masiero mechanism, but also can lead to
many variation of the  minimal scenario with interesting consequences for
ongoing and future SUSY and dark matter searches.

The LSP in this model is the lightest neutralino (in a limited region of parameter space) which is a
sizable mixture of bino, wino and higgsino, or gravitino
(in most region of parameter space) with sneutrino or stop NLSP.
Stau NLSP might be detectable using stopper and similarly for stop NLSP,
but a detailed study is
needed.
In the case of stop NLSP it is important to clarify 
whether such a scenario is consistent with BBN.

The model predicts light stop, $m_{{\tilde t}_1} \gsim 150$ GeV, and light gluino,
$m_{\tilde g} \gsim 400$ GeV, in the least fine
tuned region of parameter space. Light gluino should be easy to see at the LHC
or even at the Tevatron. In spite of stop being considerably lighter than other squarks,
it might be easier to search for the first two generations of squarks at the Tevatron.
Indeed, recent results from D0 and CDF collaborations~\cite{Abazov:2006bj,CDF_squarks+gluino},
for jets + missing transverse energy search,
put strong constraints on masses of the first two generations of squarks and the gluino mass,
in the range
$\sim 300 \, - \, 400$ GeV. These limits will further improve in near future.
At this point we would like to note
that the results of both collaborations are presented in $m_{\tilde q}$ -- $m_{\tilde g}$ plane
for mSUGRA scenario and exclusion limits cover only
gluino masses little larger than squark masses because otherwise there is no
mSUGRA solution.
However, squarks quite lighter than gluino are well motivated
by natural EWSB. In the model presented here masses of gluino and the first two generations of squarks
lie very near the border with no mSUGRA solution, and, as we discussed, models that would further
improve on naturalness (with more negative stop masses squared at the GUT scale) would lead to squarks
even lighter compared to gluino.
Therefore we strongly encourage both D0 and CDF collaborations to extend the search
and explore full kinematically allowed region in
the squark-gluino plane so that also these scenarios are covered in addition to the not so
natural one.

Considerations of natural EWSB in MSSM together with the current experimental limits on SUSY and Higgs
spectra lead us to conclusions that SUSY spectrum might be quite strange and perhaps complicated
(not unifying at any scale) compared to the usual
expectations based on models like mSUGRA, and that
 there is a good chance
we live in a meta-stable vacuum. However, as we showed, these
seemingly unattractive features might be a consequence of the same
elegant idea that leads to an understanding of quantum assignments
of standard model particles and gauge coupling unification.

\section*{Acknowledgement}

We thank K. Agashe, K. Choi, J. Ellis, G. Giudice, A. Kusenko, K.-I.
Izawa, K.-I. Okumura, M. Peskin, S. Raby and N. Weiner for
discussions. HK thanks the Galileo Galilei Institute for Theoretical
Physics and CERN for hospitality and the INFN for partial support
during the work. IK thanks LBNL for hospitality during his visit. RD
is supported in part by U.S. Department of Energy, grant number
DE-FG02-90ER40542. HK is supported by the ABRL Grant No.
R14-2003-012-01001-0, the BK21 program of Ministry of Education,
Korea and the Science Research Center Program of the Korea Science
and Engineering Foundation through the Center for Quantum Spacetime
(CQUeST) of Sogang University with grant number R11-2005-021.

\newpage

\appendix

\noindent
{\Large \bf Appendix}

\section{Calculation of supersymmetry breaking parameters at the GUT
scale \label{sec:softterms}}

We closely follow the approach and notation given in
\cite{Giudice:1997ni} \cite{Arkani-Hamed:1998kj}.
The idea is to treat couplings (gauge, Yukawa,
wavefunction renormalization) as superfields
whose scalar components are the couplings
and F components are the gaugino masses. The outcome is that
we can extract renormalization group properties of supersymmetry
breaking parameters from renormalization group equations of
ordinary couplings. It simplifies the calculation of soft supersymmetry breaking
parameters.

The running of gauge couplings at one loop is given by \bea \f{d
\a_i^{-1}}{d \log \mu} & = & \f{b_i}{2\pi}, \eea where
$b_i=(3,-1,-\f{33}{5})$ for the three gauge couplings of MSSM and
$\mu$ is the renormalization group scale. Wavefunction
renormalization ($Z_Q$) of a chiral superfield $Q$ is given by
anomalous dimensions, \bea \log Z_Q (\mu) & = & \int_{\L_{UV}}^{\mu}
\f{d\mu^{\p}}{\mu^{\p}} \g_Q (\mu^{\p}) = \sum_i \f{c_i}{\pi}
\int_{\L_{UV}}^{\mu} \f{d\mu^{\p}}{\mu^{\p}} \a_i, \eea where \bea
\g_Q & = &  \f{d \log Z_Q}{d \log \mu} = \sum_i \f{c_i}{\pi} \a_i,
\eea with $c_i$, the quadratic casimir. It can be rewritten as \bea
\log Z_Q(\mu) & = & \log Z_Q(\L_{UV}) + \sum_i \f{2c_i}{b_i}
\log \f{\a_i(\L_{UV})}{\a_i(\mu)}, \\
Z_Q (\mu) & = & Z_Q (\L_{UV}) \prod_i \left(
\f{\a_i(\L_{UV})}{\a_i(\mu)} \right)^{\f{2c_i}{b_i}}. \eea

Suppose that there is an adjoint chiral superfield $\Sigma$ which
breaks SU(5) down to the standard model gauge group. At high energy,
the beta function coefficient of the GUT group is given as $b_G =
3\times5 - 5 -3\times2 -1 = 3$ for SU(5). Each term represents the
contribution from vector supermultiplet of $SU(5)$, the adjoint
chiral multiplet of $SU(5)$, three generation of matter fields and
Higgs fields respectively. At $M_{\rm GUT}$, $X,Y$ gauge bosons
become massive by eating wouldbe Goldstone bosons in $\Sigma$. Let
us define $b_F$ as the beta function coefficient excluding X, Y
gauge bosons and $b_H$ as the one for the low energy theory. Gauge
messengers give \bea b_G - b_{Fi} & = & 3(N_C - N_{Ci}) - (N_C -
N_{Ci}), \eea which is $(4,6,10)$ for $i=3,2,1$ gauge group
respectively. There still remain (diagonal) adjoints of $\Sigma$
under the low energy gauge group which we call $\Sigma_3$ and
$\Sigma_2$ given by \bea b_{Fi} - b_{Hi} & = & -N_{Ci}, \eea which
is $(-3,-2,0)$ respectively. We call $b_{Xi} = b_G - b_{Fi}$ as the
beta function coefficient coming from fields that become massive by
$\Sigma$.
\bea b_F & = & b_{M_a} + b_H \nn \\
b_G & = & b_X + b_{M_a}+ b_H . \eea

At low energy, the degrees of freedom would be the usual gauge
bosons (or vector multiplets) of 3,2,1 and matter and Higgs fields.
\bea b_{Hi} & = & 3N_{Ci} -7, \eea which is $(2,-1,-7)$
respectively. We assume that the Higgs triplet mass is just below
the GUT scale to simplify the discussion.\footnote{In case when
Higgs triplet is heavier than the GUT scale, the final expression
becomes slightly complicated since it can not be written in terms of
single parameter $b_G$. As Higgs triplet contribution does not make
a significant change in the result, we take the simplest case (Higgs
triplet slightly lighter than the GUT scale).} The expression for
the running of a gauge coupling is then written as follows: \bea
4\pi \a^{-1} (\mu) & = & 4\pi \a^{-1} (\L) + b_X \log
\f{{\Sigma}^{\d} \Sigma}{\L^2} + b_{M_a} \log \f{M_a^2}{\L^2} + b_H
\log \f{\mu^2}{\L^2}. \eea

Gaugino masses at the messenger scale are determined by analytic
continuation of gauge couplings into superspace. \bea M_i & = &
-b_{X_i} \f{\a_i}{4\pi} \left| \f{F}{\Sigma} \right| = -b_X M_{\rm
SUSY}, \eea where $b_{X_i}$ is the contribution of fields which
become massive by $\Sigma$ and $M_{\rm SUSY}$ is defined in Eq.
(\ref{eq:MSUSY}). If there are gauge messengers and matter
messengers at the same time, $b_{X_i} = 2(5-N_{C_i})$ for massive X
and Y superfields and $b_{X_i} = -1$ for $5$ and $\bar{5}$
messengers. The explicit expression for the gauge messenger
contribution with $N_{\rm mess}$ matter messengers is \bea M_i & = &
\left[ -2(5-N_{C_i}) +N_{\rm mess} \right] M_{\rm
SUSY}.\label{eq:gaugino}  \eea

For soft scalar masses, we consider the case in which
$M_a$ is slightly lower than the messenger scale,
$\Lambda \ge \Sigma \ge M_a \ge
\mu$ so that we can write
\bea \log Z_Q (\Sigma,{\Sigma}^{\d},\mu) & = & \log Z_Q (\L) +
\f{2 c_G}{b_G} \log \f{\a_G(\L)}{\a_G(\Sigma)} \nn \\ && + \sum_i
\f{2c_i}{b_{M_ai}+b_{H_i}} \log \f{\a_i(\Sigma)}{\a_i(M_a)} + \sum_i
\f{2c_i}{b_{H_i}} \log \f{\a_i(M_a)}{\a_i(\mu)}. \eea We can assume
that the scale difference between $M_a$ and $\la \Sigma \ra$ is
negligible. With $\xi_i = \f{\a_i (\Sigma)}{\a_i (\mu)}$, the same
calculation as in the previous subsection gives \bea m_Q^2 =
\left(2c_G  b_G - \sum_i \f{2c_i}{b_{M_ai} + b_{H_i}} b_{G}^2 +
\sum_i (\f{2c_i}{b_{M_ai} + b_{Hi}}- \f{2c_i}{b_{H_i}}) b_{Xi}^2 +
\sum_i \f{2c_i}{b_{H_i}} \f{b_{Xi}^2}{\xi_i^2} \right) M_{\rm
SUSY}^2 \nn \eea At $\mu = M_a \sim \la \Sigma \ra$, we obtain soft
scalar masses, \bea m_Q^2 & = &\left(2c_G b_G + \sum_i
\f{2c_i}{b_{M_ai} + b_{H_i}} (-b_{Gi}^2 + b_{Xi}^2)  \right)
M_{\rm SUSY}^2 \nn \\
&& = \left( 2\Delta c  b_G - 2\sum_i c_i b_{Xi}\right) M_{\rm
SUSY}^2, \label{eq:scalars} \eea where $\Delta c = c_5 - \sum_i
c_i$. For the minimal content ($V$, $\Sigma$, Higgs and matter
fields), we have $b_G =3$. By adding one extra $5+\bar{5}$
messenger, $b_G$ is lowered by one.

The $A$ terms at the messenger scale are calculated by canonically
normalizing the scalar fields, \bea A_i (M) & = & \left. \f{\partial
\log Z_{Q_i} (\Sigma,{\Sigma}^{\dagger},\mu)}{\partial \log \Sigma}
\right|_{\Sigma=M} \f{F}{M}. \eea In the gauge messenger model, \bea
A_Q (M) & = & 2 \Delta c_{Q} M_{\rm SUSY} \label{eq:aterm}, \eea and
similarly for others. From the Table \ref{tab:casimir} we see
$2\Delta c_{Q} = 3$, $2\Delta c_{H_u} = 3$ and $2\Delta c_{u^c} =
4$. The $A$ term for top Yukawa coupling is then \bea A_t (M) & = &
A_Q + A_{H_u} + A_{u^c} = 10 M_{\rm SUSY}. \eea

Quadratic casimirs and related parameters (e.g., $\Delta c$, $\sum_i
c_i b_{Xi}$) used in the calculations are summarized in
Table~\ref{tab:casimir}.

\begin{table}[t]
\bea \ba{|c|ccccc|} \hline
& ~~~Q~~~ & ~~~u^c~~~ & ~~~d^c~~~ & ~~~L~~~ & ~~~e^c~~~ \\
\hline
c_3 & \f{4}{3} & \f{4}{3} & \f{4}{3} & 0 & 0 \\
&&&&& \\
c_2 & \f{3}{4} & 0 & 0 & \f{3}{4} & 0 \\
&&&&& \\
c_1 & \f{1}{60} & \f{4}{15} & \f{1}{15} & \f{3}{20} & \f{3}{5} \\
&&&&& \\
c_5 & \f{18}{5} & \f{18}{5} & \f{12}{5} & \f{12}{5} & \f{18}{5} \\
&&&&& \\
c_{10} & \f{45}{8} & \f{45}{8} & \f{45}{8} & \f{45}{8} & \f{45}{8}
\\
&&&&& \\
\Delta c  & \f{3}{2} & 2 & 1 & \f{3}{2} & 3 \\
&&&&& \\
-2\sum_i c_i b_{Xi} & -20 & -16 & -12 & -12 & -12 \\
\hline \ea \nn \eea \caption{ \label{tab:casimir} Quadratic casimirs
and their combinations relevant for calculation of soft SUSY
breaking masses. $c_5$ is the quadratic casimir under $SU(5)$ and
$c_{10}$ is the one under $SO(10)$. In the final expression, the
minimal messenger model, $b_X = (4,6,10)$, has been used.}
\end{table}

\section{Suppression of Gravity Mediation \label{sec:gravity}}

\subsection{Large cutoff scenario}

Gravity mediated contribution can not be neglected in gauge
messenger model due to high messenger scale $M_{\rm GUT}$ ($
\f{m_{3/2}}{M_{\rm SUSY}} \simeq 1.5$). The problem can be overcome
either by raising up the cutoff scale of the theory beyond the
Planck scale or lowering the messenger scale (GUT scale). There
would be various ways of achieving it and here we illustrate some
possibilities. We consider superconformal frame and Einstein frame
to discuss the problem. Fine tuning of electroweak symmetry breaking
is not sensitive to the choice of frames but neutralino LSP (or
NLSP) region can be enlarged in Einstein frame.

\begin{itemize}

\item Sequestering (Large cutoff in superconformal frame)

Conformal symmetry guarantees the stability of the sequestering
once it happens at tree level.
\bea
S_{\rm SUGRA} & = & \int d^4 x \left[ \int d^4 \theta
 {\mathbb E} \left( -3 M_{\rm Pl}^2
e^{-\f{K}{3 M_{\rm Pl}^2}} \right) + \left\{ \int d^2 \theta \left(
\frac{1}{4} f_a W^{a\alpha} W_{a \alpha} + W  \right) + {\rm h.c.}
\right\} \right], \nn \eea If K\"ahler potential is minimal in the
superconformal frame, \bea -3 M_{\rm Pl}^2 e^{-\f{K}{3 M_{\rm
Pl}^2}} & = & \Phi^{\dagger} \Phi + \Sigma^{\dagger} \Sigma, \nn
\eea there would be no dangerous gravity mediation effect. We do not
address how the conformal sequestering can be realized in our
specific setup. Sequestered form of K\"ahler potential is understood
by conformal symmetry. Conformal symmetry prevents higher
dimensional operators. Sequestering means a large cutoff for
possible non-renormalizable interactions with order one
coefficients. \bea -3M_{\rm Pl}^2 e^{-\f{K}{3 M_{\rm Pl}^2}} & = &
\Phi^{\dagger} \Phi + \Sigma^{\dagger} \Sigma + \f{1}{M_*^2}
\Sigma^{\dagger} \Sigma \Phi^{\dagger} \Phi + \cdots, \eea with $M_*
\gg M_{\rm Pl}$. Note that $M_* \sim 5 M_{\rm Pl}$ is enough to keep
an accuracy of gauge messenger model within 1 or 2 percent. Required
suppression is very small and slightly large cutoff might work
without building a sequestering model.\footnote{Gauginos can get a
correction 10 to 15 percent in this case but this contribution does not lead to
flavor changing neutral currents.}

\item Large cutoff in Einstein frame

Universal soft scalar masses appear for
the minimal K\"ahler potential in Einstein frame,
$K = \Phi^{\dagger} \Phi +
\Sigma^{\dagger} \Sigma$. \bea -3M_{\rm Pl}^2 e^{-\f{K}{3 M_{\rm
Pl}^2}} & = &
-3M_{\rm Pl}^2  + K -\f{1}{6 M_{\rm Pl}^2} K^2 + \cdots, \nn \\
& = &  \Phi^{\dagger} \Phi + \Sigma^{\dagger} \Sigma - \f{1}{3
M_{\rm Pl}^2} \Sigma^{\dagger} \Sigma \Phi^{\dagger} \Phi + \cdots.
\nn \eea The last term gives universal soft scalar masses to all
$\Phi$s once $F_\Sigma$ is nonzero\footnote{When there are several
sources of supersymmetry breaking, all of them contribute to
$m_{3/2}$.} and we have $\delta V = m_{3/2}^2 \Phi^{\dagger} \Phi$.
The problem associated with other unpredictable soft terms due to
nonrenormalizable operators can be solved if a large cutoff of the
theory is assumed \cite{Ibe:2004mp} \cite{Ibe:2006fs}. Let the
cutoff of the theory be $M_*$. We can imagine that matter sector
couples weakly while gravity sector happens to couple strongly.

\end{itemize}

There are two ways to explain large cutoff. Firstly, we can start
with the cutoff $M_*$ and the observed Planck scale happens to be
small due to the cancelation with loop corrections $\delta M_*$. $S
= \int d^4 x (M_*^2 + \delta M_*^2) R = \int d^4 x M_{\rm Pl}^2 R +
\cdots .$ Numerically $M_{\rm Pl} \sim \f{M_*}{4\pi} \sim
\f{M_*}{10}$. The observed Planck scale appears to be lower than the
cutoff of the theory due to an accidental cancelation of the bare
parameter and the quantum corrections. The other explanation comes
with a strong coupling. \bea S & = & \int d^4 x \f{1}{g^2} \left[
M_*^2 R + \cdots \right]. \eea If the theory couples strongly, $g
\sim 4\pi$, we would get an effective Planck scale $M_{\rm Pl} =
\f{M_*}{g} \sim \f{M_*}{4\pi}$. Now if $M_* \sim 3.0 \times 10^{19}$
GeV, we would get the reduced Planck scale $M_{\rm Pl} = 2.4 \times
10^{18}$ GeV. $M_*$ is much larger than $M_{\rm Pl}$. It is natural
to have a reduced Planck scale if the gravity couples strongly.

Similarly we can imagine that each sector can couple with a
different strength. Naive dimensional analysis
\cite{Weinberg:1978kz} \cite{Luty:1997fk} \cite{Cohen:1997rt}
 tells us that \bea K
& = & \f{M_*^2}{g^2} \hat{K} ( \f{g\Sigma}{M_*},
\f{g\Sigma^{\dagger}}{M_*}) \eea where $\Sigma$ couples strongly
with $g \sim 4\pi$. When there is a weakly coupled sector, we can
add them to the K\"ahler potential as follows. \bea K & = &
\f{M_*^2}{g^2} \hat{K}_1 ( \f{g\Sigma}{M_*},
\f{g\Sigma^{\dagger}}{M_*}, \f{e\Phi}{M_*}, \f{e\Phi^{\dagger}}{M_*}
)  + \f{M_*^2}{e^2} \hat{K}_2 ( \f{e\Phi}{M_*},
\f{e\Phi^{\dagger}}{M_*}), \eea where $\Phi$ represents all fields
that couple weakly by itself with $e \sim 1$ and $\hat{K}$ has
polynomials with order one coefficients. Expanding K\"ahler
potential up to quartic terms, we get \bea K & = & \Sigma^{\dagger}
\Sigma + \Phi^{\dagger} \Phi + \f{g^2}{M_*^2} (\Sigma^{\dagger}
\Sigma)^2 + \f{e^2}{M_*^2} (\Phi^{\dagger} \Phi)^2 + \f{e^2}{M_*^2}
\Sigma^{\dagger} \Sigma \Phi^{\dagger} \Phi. \eea Note that $M_{\rm
Pl} \sim \f{M_*}{g} \sim \f{M_*}{4\pi}$. We can consider the case in
which matter fields $\Phi$ (quarks and leptons) couple weakly ($e
\sim 1$) while Higgs fields $H$ and $\bar{H}$ couple strongly ($g
\sim 4\pi$). Relevant terms in the K\"ahler potential would be \bea
K & = & \Sigma^{\dagger} \Sigma + H^{\dagger} H +{\bar{H}}^{\dagger}
\bar{H} + \Phi^{\dagger} \Phi \nn
\\
&& + \f{1}{M_{\rm Pl}^2} \Sigma^{\dagger} \Sigma H^{\dagger} H +
\f{1}{M_{\rm Pl}^2} \Sigma^{\dagger} \Sigma {\bar{H}}^{\dagger}
\bar{H} + \f{1}{M_*^2} \Sigma^{\dagger} \Sigma \Phi^{\dagger} \Phi.
\eea Giudice-Masiero term \bea K & = & \f{1}{M_{\rm Pl}} H
\Sigma^{\dagger} \bar{H} + \f{1}{M_{\rm Pl}^2} \Sigma^{\dagger}
\Sigma \bar{H} H, \eea is suppressed only by $M_{\rm Pl}$. This
setup explains $\mu \sim m_{3/2}$ and $B\mu \sim m_{3/2}^2$. Let us
summarize gravity mediated contributions on various fields when only
Higgs fields couple strongly at $M_{\rm Pl}$ and the large cutoff is
realized for other fields in superconformal frame. \bea
&& m_{H_u}^2, ~~ m_{H_d}^2, ~~ \mu^2, ~~ B\mu \sim m_{3/2}^2, \nn \\
&& m^2 ({\rm squarks, sleptons}) \sim \f{m_{3/2}^2}{16\pi^2}, \nn \\
&& M_{\f{1}{2}}, ~~ A \sim \f{m_{3/2}}{4\pi}. \nn \eea Note that
gravity mediation is suppressed in squark and slepton soft scalar
masses and gaugino masses. In Einstein frame, a common universal
$m_{3/2}^2$ is added to all squarks, sleptons and Higgs soft scalar
masses.

\subsection{Lowering the GUT scale}

Another way of suppressing gravity mediation is to lower the GUT
scale. Although we have an indirect evidence that three gauge
couplings meet at the GUT scale, $M_{\rm GUT}  = 2 \times 10^{16}
\rm GeV$, any hints of $X$ and $Y$ gauge bosons have not been
observed yet. The lower bound on their mass due to proton decay from
dimension six operators is about $10^{15}$ GeV. If we can suppress
the proton decay from dimension five operators related to color
triplet Higgses, we can lower the GUT scale (more precisely the
messenger scale, the mass of X, Y gauge bosons and X, Y gauginos).
GUT scale threshold correction would then explain the illusion of
having $M_{\rm GUT} = 2 \times 10^{16}$ GeV. Furthermore, by adding
extra matter fields in full multiplets of $SU(5)$, we can also make
$\alpha_{\rm GUT}$ larger than $1/24$ while keeping unification.
This would enhance the gauge mediation effects even for messenger
scale being the GUT scale. Finally, a combination of both effects
might suppress gravity contribution to a negligible level.



\end{document}